\documentclass[a4paper]{article}

\usepackage[english]{babel}
\usepackage[utf8x]{inputenc}
\usepackage[T1]{fontenc}

\usepackage[a4paper,top=3cm,bottom=2cm,left=3cm,right=3cm,marginparwidth=1.75cm]{geometry}

\usepackage{amsmath}
\usepackage{graphicx}
\usepackage[colorinlistoftodos]{todonotes}
\usepackage[colorlinks=true, allcolors=blue]{hyperref}

\title{DMTCP Checkpoint/Restart of MPI Programs via Proxies}
\author{Gregory Price}

\begin{document}
\maketitle

\begin{abstract}
MPI accomplishes portable, standardized message-passing between processes by exposing a standard API that hides the implementation of the underlying mechanism for message passing.  Until now, checkpointing an MPI program required knowledge of these underlying mechanisms.  Through the addition of a proxy, we demonstrate that MPI programs can be checkpointed and restarted regardless of the MPI implementation utilized.  Further, proxies may enable MPI programs to be checkpointed on one MPI implementation, and restarted on another.
\end{abstract}

\section{Introduction}

This work has largely been produced using examples from other DMTCP plugins from the core dmtcp library \cite{dmtcp}.  It was implemented and tested against MPICH \cite{mpich}, with examples from mpitutorial.com \cite{mpitut}.  

MPI provides the HPC community a standardized way to pass messages between processes in parallel computation.  Each MPI implementation provides an opaque, standardized interface to send and receive messages; however, each
implementation may differ in system requirements and mechanisms used to actually pass the message (i.e., sockets, pipes, shared memory, etc.).  

This provides a challenge when checkpoint/restart is desired.  Classically, checkpoint/restart is achieved via virtualization of various resources (such as file descriptors, sockets, shared memory, etc.), which are used transparently by the user application.  Checkpoint/restart software must then provide support for each potential implementation of MPI;  agnosticism is a pipe dream in this reality.

Through the use of MPI proxies and a DMTCP Plugin, we present a solution that provides simple, portable, transparent, and agnostic Checkpoint/Restart.  Our implementation requires no changes to application code, and can be deployed on any architecture machine that supports DMTCP.

Our strategy makes one, key, novel observation about how to checkpoint MPI: don't.  By interposing on the MPI library, and moving all MPI procedures into a separate process, we only need to checkpoint a single, ephemeral interface between the application program and the proxy.  Then, we simply, transparently translate MPI calls into calls to our MPI proxy.

Restart presents a separate problem, one which is fairly understood within the realm of checkpoint/restart:  How do we handle messages in flight?  There are two options for managing the situation.

\begin{enumerate}
\item Log and replay of messages on restart
\item Drain the network of in-flight messages, and cache them for restart
\end{enumerate}

Logging many messages requires additional (potentially significant) overhead throughout the lifetime of the computation, while draining
the network only incurs a cost at the time of checkpoint.  We employ network draining as our mechanism of choice, under the assumption that a one-time cost during checkpoint is more tolerable to HPC professionals. This overhead can also be easily controlled through changing how often a checkpoint is created.

Finally, we believe solving this checkpoint/restart issue for MPI via the combination of library interposition (via DMTCP plugins) and a proxy provides a compelling case for using this strategy elsewhere.  Other technologies that provide a standardized interface to an external service can also benefit from checkpoint/restart via the use of proxies.

\section{Background}
MPI provides two main services: 
\begin{enumerate}
\item Creation of processes (called ranks) for distributed computation
\item Message passing between these ranks
\end{enumerate}

\begin{figure}[h!t]
\begin{center}
\includegraphics[width=0.70\textwidth]{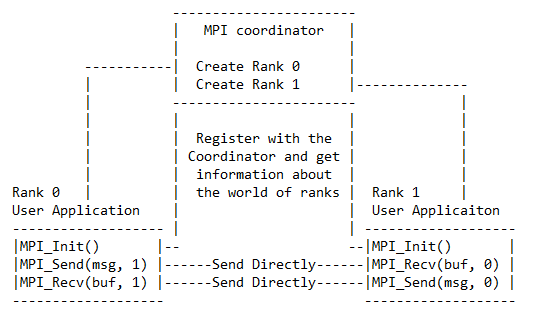}
\caption{\label{fig:figure1}The communication structure of a simple MPI application}
\end{center}
\end{figure}

A coordinator is used for exchanging information about rank identification, how to build direct communication paths between ranks, and what groups have been created for publish-subscribe style communications (Broadcasting to a subset of ranks).  It is useful to visualize this relationship in Figure \ref{fig:figure1}, as one possible implementation of an MPI coordinator.

From here, we can see that there is a parent-child process relationship.  The Coordinator is a parent to both Rank 0 and Rank 1. Also, there are at least three ``communication interfaces'' created between these three programs: Interfaces between each rank and the coordinator, and between each rank.  

Each rank has a separate ``MPI Library'', which is used by the application program to interact with the coordinator and other ranks.  We call this an \textit{Active MPI Library}. An ``Active'' library is one whose functions are loaded and executed, whereas a ``passive'' library is one whose functions may be referenced but never used.

\begin{figure}[h!t]
\begin{center}
\includegraphics[width=0.6\textwidth]{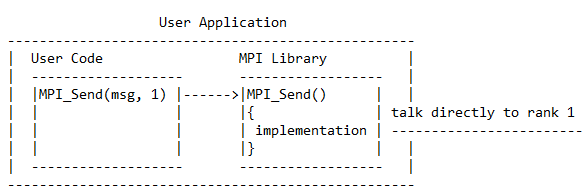}
\caption{\label{fig:figure3}MPI library in User Application }
\end{center}
\end{figure}

Prior work on checkpointing MPI programs had been built on the assumption that all external resources within the entire system must be virtualized.  We can visualize this with a ``checkpoint boundary'' for an arbitrary MPI computation.

\begin{figure}[h!t]
\begin{center}
\includegraphics[width=0.75\textwidth]{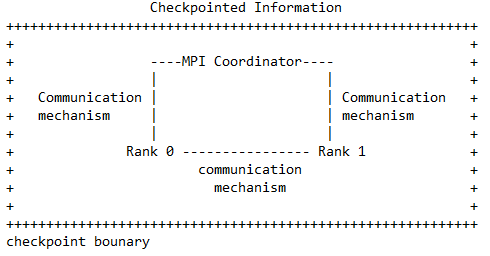}
\caption{\label{fig:figure2}The checkpoint boundary of existing solutions}
\end{center}
\end{figure}

Each relationship and communication mechanism within the checkpoint boundary will increase the complexity of a successful checkpoint/restart approach as the number of potential communication mechanisms used by MPI
implementations increases.  (The ``checkpoint boundary'' encapsulates all the resources that must be recorded and/or restored upon restart.)

\section{Our Approach}

We seek to avoid checkpointing any implementation-specific MPI code, and instead create a network with a checkpoint boundary that looks as follows:

\begin{figure}[h!t]
\begin{center}
\includegraphics[width=0.75\textwidth]{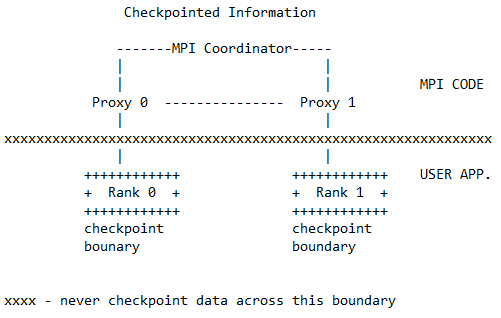}
\caption{\label{fig:figure4}Our Checkpoint Boundary}
\end{center}
\end{figure}

This reduces the total number of interfaces needed to be virtualized, and completely segregates the MPI implementation from the user application.  The MPI library used by the user application is no longer ``active'', and is replaced by a DMTCP plugin that provides the virtual interface to the ``active'' library made available by the proxy.

\section{challenges}

By introducing an external proxy, a one-step message-passing mechanism is broken into two steps.  This introduces a temporal issue that was not present in other checkpointing solutions.  We must deal with two
challenges that must be solved in order to successfully checkpoint and restart:

\begin{enumerate}
\item In-flight Data.  Rank 0 sends Rank 1 a message, the message has been received by Proxy 0, but has not been transferred to Proxy 1 or Rank 1. A checkpoint is initiated, leaving the message stuck ``in flight''.
\item Proxy State.  Rank 0 makes a call to receive a message, but Rank 1 has not sent the message yet.  Rank~0 must re-issue the call in order to receive on restart.
\end{enumerate}

In both cases, we recognize that a potential for failure occurs when the proxy has a ``state'' that is not being recorded due to our chosen checkpoint boundary. We revert to the two basic techniques discussed previously (Drain, and Log-and-Replay) to solve these in the simplest way possible.

For ``in-flight data'', we drain the network of all in-flight messages, and cache this information in the receiving processes' memory space.  In order to know when all in-flight data has been successfully drained, we utilize the DMTCP coordinator to share the number of messages that each rank has sent and received. When draining, we continue to increment these numbers.  When the numbers are equivalent, we know that there is no longer any data in flight, and thus we can move forward with checkpointing.  This heuristic was previously used in the PhD thesis of Jiajun Cao for checkpointing over InfiniBand \cite{jiajun}.

For Proxy-State replication, we break messages into two basic categories:  
\begin{enumerate}
	\item Administrative messages
	\item Message actions
\end{enumerate}

Administrative messages are messages between the rank and the MPI coordinator to either retrieve information about the current configuration of the MPI ranks, or to create new configurations (such as new groups of ranks). Configuration messages must be replayed to ensure that the ``Active'' MPI library in the proxy is in the same state as when we initiated the checkpoint.

Message actions are calls such as MPI\_IProbe and MPI\_Recv, where the user application is trying to retrieve either data, or information about the state of the  network. For example, MPI\_IProbe is used to check if there is a message waiting to be received.  The relevant information for these message actions must be cached at checkpoint time.  The simplest example is the drain mechanism for in-flight data --- these messages are cached for use on restart.  A call to MPI\_Recv must check the cache for a buffered message before checking the proxy for a buffered message. 

\section{API Support}

Currently only the following calls to the MPI interface are fully supported.
\begin{itemize}
\item MPI\_Init
\item MPI\_Finalize
\item MPI\_Comm\_size
\item MPI\_Comm\_rank
\item MPI\_Type\_size
\item MPI\_Send
\item MPI\_Recv
\item MPI\_Probe
\item MPI\_Iprobe
\item MPI\_Get\_count
\end{itemize}

Future work will expand support to the following API calls:
\begin{itemize}
\item Non-blocking Asynchronous Send / Receive
\begin{itemize}
\item MPI\_Isend
\item MPI\_Irecv
\item MPI\_Test
\end{itemize}

\item Collective Communication
\begin{itemize}
\item MPI\_Bcast
\item MPI\_Barrier
\item MPI\_Scatter
\item MPI\_Gather
\item MPI\_Allgather
\item MPI\_Reduce
\item MPI\_Allreduce
\end{itemize}

\item Communicators and Groups
\begin{itemize}
\item MPI\_Comm\_group
\item MPI\_Group\_incl
\item MPI\_Comm\_create\_group
\item MPI\_Group\_free
\item MPI\_Comm\_free
\end{itemize}
\end{itemize}

\section{Current State}

Presently we support checkpointing of simple Send/Receive programs that use any of the supported API calls.  We do not currently support asynchronous send and receive (i.e., non-blocking sends and receives). Trivially, we support non-blocking receives by supporting MPI\_Iprobe, which allows user applications to determine if a message is waiting to be received.  However, support of MPI\_Isend will require the caching of additional data (not yet implemented).

With these basic building blocks supported, we can confidently say that the remaining API calls can also be supported --- it is now a simple matter of plumbing, and no longer a question of feasibility.

\section{Future Work}

Along with adding full API support, we also intend to explore the idea of checkpointing an application on one MPI implementation, and then restarting the application on a different MPI implementation.  This will be accomplished via two core mechanisms.  First, we will provide a virtualized interface to MPI specific resources, such as an MPI\_Request id or MPI\_Status structure.  Second, we will provide a new layer to handle translations between the \textit{Passive MPI Library} in the MPI Plugin and the \textit{Active MPI Library} in the MPI Proxy.  We believe this approach will make it possible to, for example, checkpoint a program built and run on MPICH, and restart the same program on an OMPI implementation without recompilation.

\section{Acknowledgements}

Gene Cooperman and Rohan Garg (Northeastern University) provided guidance on DMTCP plugin creation, and in the design of the proxy architecture.  The initial idea for the proxy architecture was developed by Rohan Garg,
Gene Cooperman and Jiajun Cao \cite{jiajun} during the development of other checkpoint/restart technologies.  This report and project reinforces their claim that a proxy architecture for checkpointing can be applicable much more generally.

The members of the High Performance Computing Lab at Northeastern University 
provided feedback and support during the development of this project.

\end{document}